\documentclass[twocolumn,showpacs,showkeys,preprintnumbers,amsmath,amssymb]{revtex4}  
\usepackage{graphicx}% Include figure files
\usepackage{dcolumn}% Align table columns on decimal point
\usepackage{bm}% bold math
\usepackage{multirow}
\usepackage{color}

\begin{document}

\title{Robust optical emission polarization in ${\mathrm{MoS}}_{2}$ monolayers\\
 through selective valley excitation}

\author{G.~Sallen$^1$}
\author{L.~Bouet$^1$}
\author{X.~Marie$^1$}
\author{G. Wang$^2$}
\author{C.R. Zhu$^2$}
\author{W.P.~Han$^3$}
\author{Y.~Lu$^3$}
\author{P.H.~Tan$^3$}
\author{T.~Amand$^1$}
\author{B.L.~Liu$^2$}
\email{blliu@iphy.ac.cn}
\author{B.~Urbaszek$^1$}
\email{urbaszek@insa-toulouse.fr}
\affiliation{%
$^1$Universit\'e de Toulouse, INSA-CNRS-UPS, LPCNO, 135 Av. de Rangueil, 31077 Toulouse, France}
\affiliation{%
$^2$Beijing National Laboratory for Condensed Matter Physics, Institute of Physics, Chinese Academy of Sciences, Beijing 100190, China}
\affiliation{%
$^3$State Key Laboratory of Superlattices and Microstructures, Institute of Semiconductors, Chinese Academy of Sciences, Beijing 100083, China}

\date{\today}

\begin{abstract}
We report polarization resolved photoluminescence from monolayer ${\mathrm{MoS}}_{2}$, a two-dimensional, non-centrosymmetric crystal with direct energy gaps at two different valleys in momentum space. The inherent chiral optical selectivity allows exciting one of these valleys and close to 90\% polarized emission at 4K is observed with 40\% polarization remaining at 300K. The high polarization degree of the emission remains unchanged in transverse magnetic fields up to 9T indicating robust, selective valley excitation.

\end{abstract}

                             % Classification Scheme
                             % 78.55.Cr   III-V semiconductors  
                             % 73.21.La    Quantum dots   
                             % 78.66.Hc  Optical properties 
                           %Use showkeys class option if keyword
\maketitle

\begin{figure}
\includegraphics[width=0.47\textwidth]{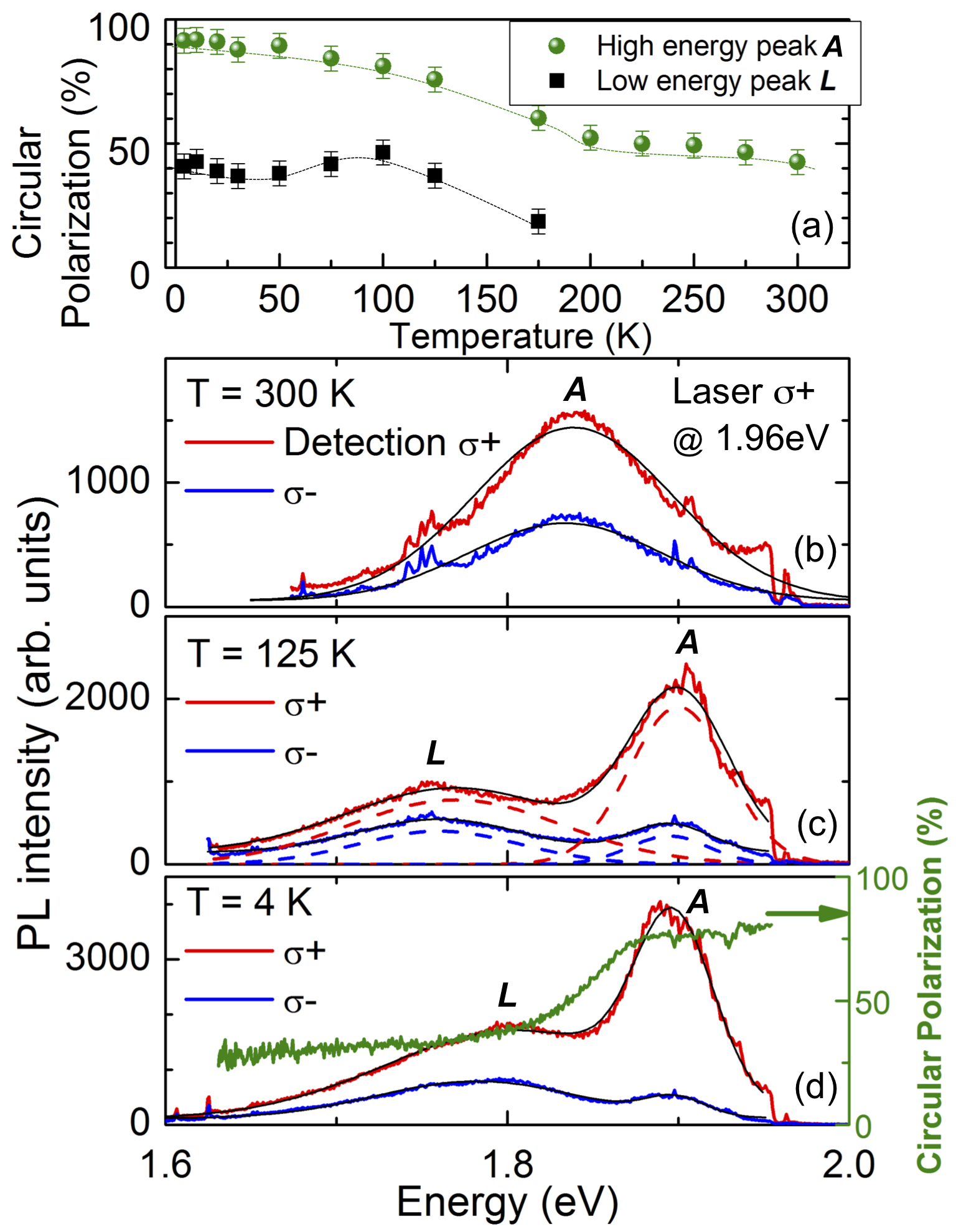}
\caption{\label{fig:fig1}(Color online) (a) Measured PL circular polarization degree of a 1ML MoS$_2$ as a function of temperature. Low energy peak (\textbf{\textit{L}}) not detectable for $T>175$~K. Dotted lines are a guide to the eye(b) PL spectra co-polarized (red/light gray) and cross-polarized (blue/dark gray) with respect to the $\sigma^+$ polarized excitation laser (HeNe at 1.96eV). At 300~K only the high energy peak \textbf{\textit{A}} is detectable. Black curves indicate fit to the data with Gaussians. (c) same as (b) but at  $T=125$~K. Both \textbf{\textit{L}} and \textbf{\textit{A}} peaks are detectable and the polarization degree is extracted by fitting them separately (dashed lines).  (d) same as (c) but at $T=4$~K, the (green) right axis shows the circular polarization. PL intensity drops as a function of temperature, counts in (b)-(d) are normalized for better visibility.
}
\end{figure}
\begin{figure}
\includegraphics[width=0.47\textwidth]{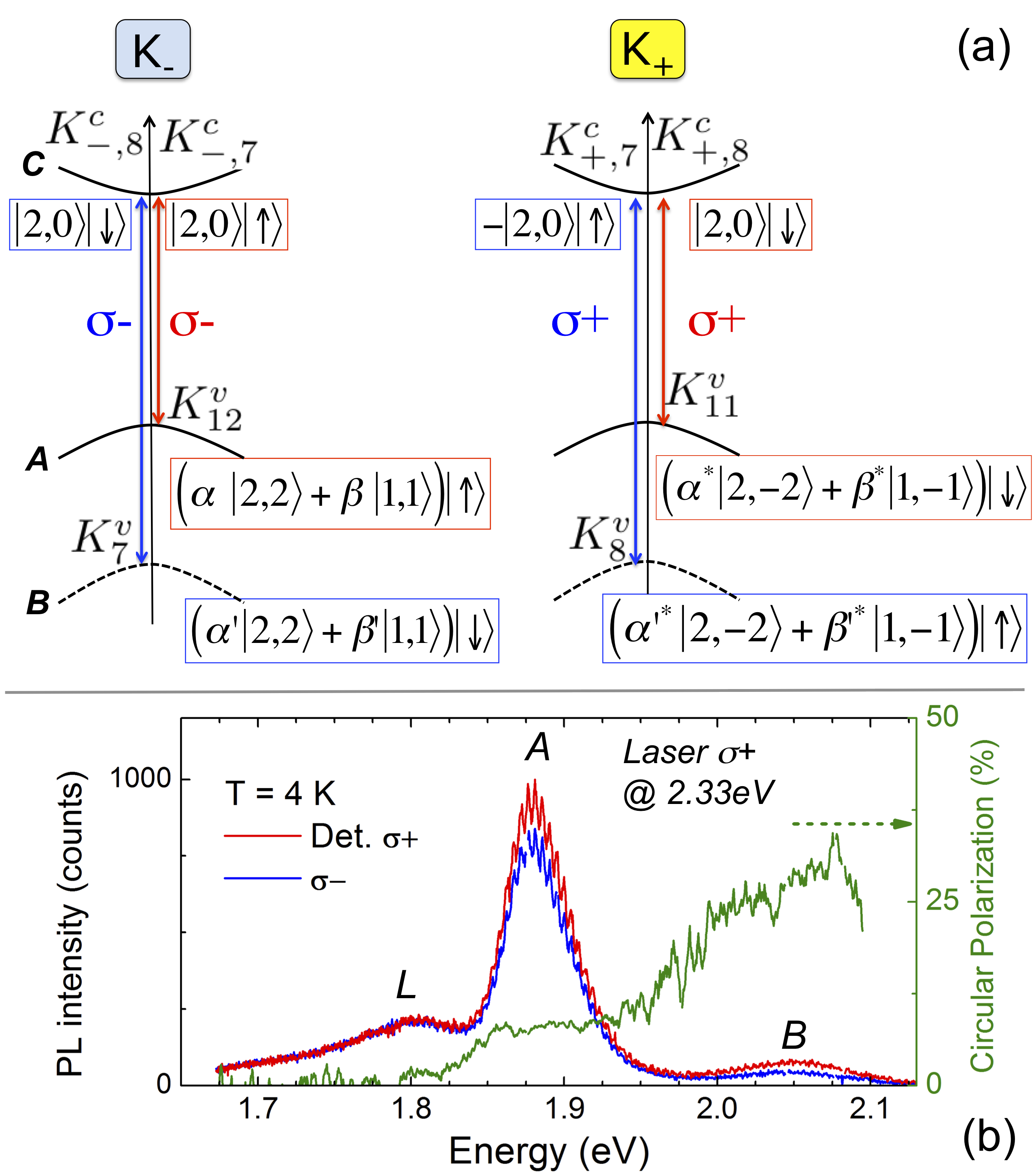}
\caption{\label{fig:fig2}(Color online) (a) Selection rules for optical dipole transitions in a simple single particle picture in $K_+$ and $K_-$ valleys at zero magnetic field, adapted from \cite{Xiao:2012a,Cao:2012a,Zhu:2011a}, states are represented as $|L,M\rangle$ for valence states, $|\uparrow\rangle$ ($|\downarrow\rangle$) stand for electron spin up (down). For simplicity, the k-linear term in the CB is neglected, the \textbf{\textit{A(B)}} VBs for $K_+$ and $K_-$ are degenerate in energy. Local trigonal symmetry of electron states at $K_{\pm}$ points is $C_{3h}$ in Koster notation \cite{Koster:1963a}, states labelled using  corresponding group representations (b). PL spectra at $T=4$~K co-polarized (red/light gray) and cross-polarized (blue/dark gray) with respect to the $\sigma^+$ polarized highly non-resonant excitation laser at 2.33eV. Transitions \textbf{\textit{L}} and \textbf{\textit{A}} are visible in addition to the \textbf{\textit{B}} exciton. The small periodic signal variations are due to optical interference in the set-up. The (green) right axis shows the PL circular polarization. 
}
\end{figure}

\begin{figure}
\includegraphics[width=0.47\textwidth]{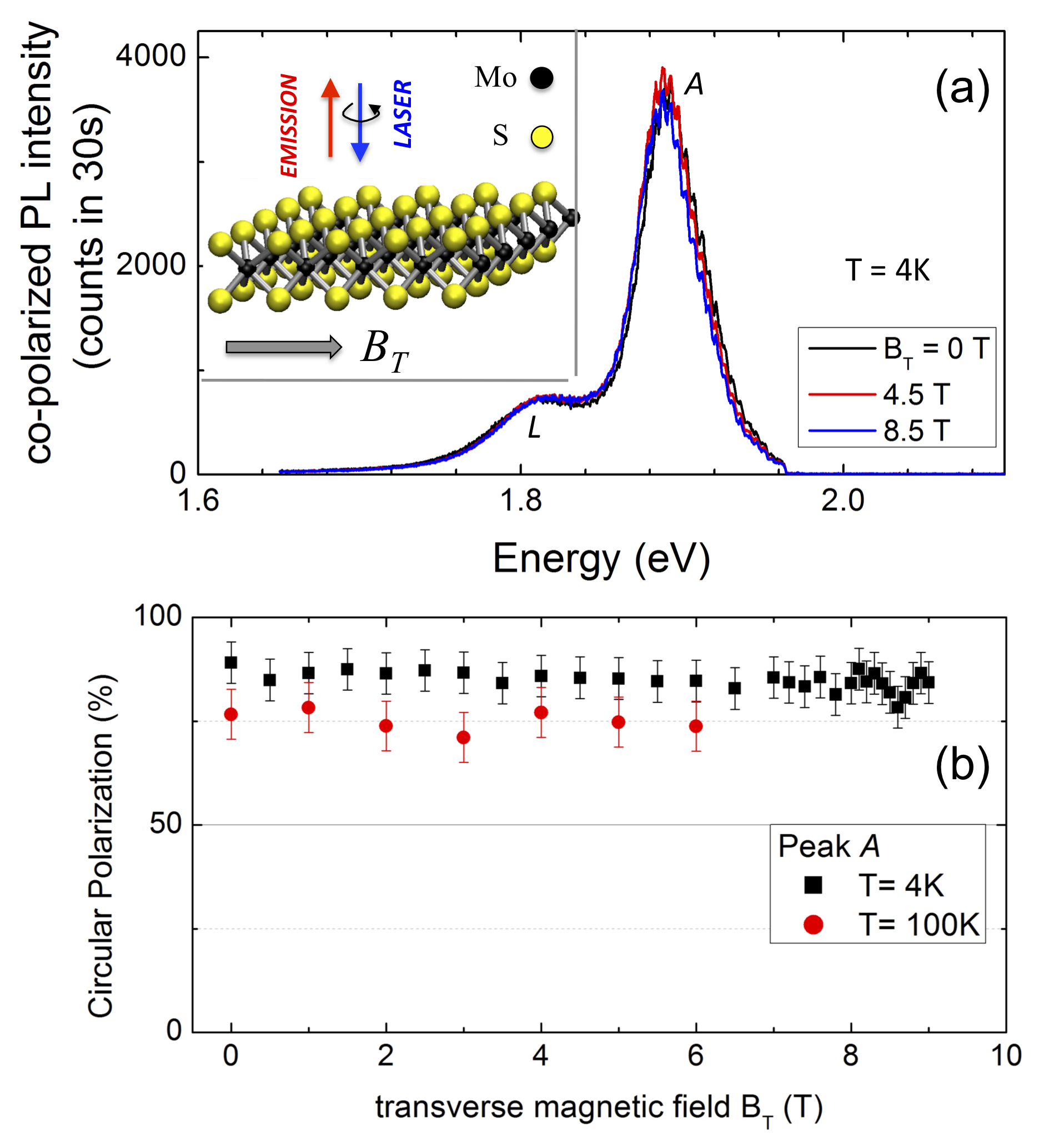}
\caption{\label{fig:fig3}(Color online) (a) Comparison of PL spectra of a single MoS$_2$ ML recorded for a transverse magnetic field $B_T=0, 4.5$ and $8.5$~T under otherwise identical experimental conditions using HeNe laser at 1.96eV. Inset: experimental geometry. (b) Measured PL circular polarization degree of the PL as a function of the applied transverse magnetic field $B_T$ for \textbf{\textit{A}} exciton at $T=4$~K (squares) and at $T=100$~K (circles). 
}
\end{figure}

 
\textit{Introduction.---}The spectacular progress in controlling the electronic properties of graphene \cite{Novoselov:2004a,Castroneto:2009a} has triggered research in alternative   atomically thin two-dimensional crystals. Here monolayer (ML) MoS$_2$ (see structure in inset of Fig.\ref{fig:fig3}a) has emerged as a very promising material for optical and spin applications for mainly two reasons. First, the indirect bulk semiconductor MoS$_2$ becomes direct when thinned to one ML \cite{Mak:2010a,Splendiani:2010a,Eda:2011a}, resulting in efficient optical absorption and emission \cite{Mak:2010a,Splendiani:2010a,Eda:2011a,Korn:2011a} and promising reports on using ML MoS$_2$ as an active field effect transistor (FET) channel \cite{Radisavljevic:2011a}. Second, inversion symmetry breaking (usually absent in graphene) together with the spin-orbit interaction leads to a coupling of carrier spin and k-space valley physics, i.e. the circular polarization ($\sigma^+$ or $\sigma^-$) of the absorbed or emitted photon can be directly associated with selective carrier excitation in one of the two non-equivalent K valleys ($K_+$ or $K_-$, respectively). This chiral optical valley selectivity has been theoretically predicted \cite{Xiao:2012a,Cao:2012a,Zhu:2011a} with very recent encouraging experimental results reporting indeed strong polarization of the photoluminescence (PL) emission in ML MoS$_2$ \cite{Mak:2012a,Zeng:2012a,Cao:2012a}.\\
\indent The selection rules for the direct, dipole allowed optical transitions in a MoS$_2$ ML  are represented in Fig.\ref{fig:fig2}a. The conduction band (CB) minimum located at the $K_-$ and $K_+$ points is spin degenerate \cite{CB}. In contrast, at the valence band (VB) maximum, also located at the $K_{\pm}$ points, spin-orbit coupling induces a splitting between spin up and spin down bands of about 150~meV due to the absence of inversion symmetry in \textit{monolayer} MoS$_2$ \cite{Xiao:2012a}. Therefore at low temperature a spin flip to the opposite spin state within the same valley is energetically forbidden i.e. a flip between VBs \textbf{\textit{A}} and \textbf{\textit{B}}, see Fig.\ref{fig:fig2}a \cite{Mak:2012a}.  Another striking feature is the reversed order of valence electron spin up and down states when going from the $K_-$ to $K_+$ valley.  So to change its spin state, the valence carrier has to (i) change valley i.e. compensate a change in wavevector comparable to the size of the Brillouin zone and (ii) in addition its orbital angular momentum (see Fig.\ref{fig:fig2}a). As a result, a change in valence carrier spin state and hence valley in k-space will be far less likely in a MoS$_2$ ML than the common spin flip mechanisms evoked in well studied semiconductors like GaAs \cite{Dyakonov:2008a}, making ML MoS$_2$ a promising system for the development of photo-induced spin Hall and valley Hall experiments.
Several original experimental schemes propose to use a stable valley index in analogy to the electron charge or spin as an information carrier in AlAs \cite{Gunawan:2006a} and graphene samples with deliberately broken inversion symmetry \cite{Zhang:2011a,Rycerz:2007a,Zhou:2007a}. The simple valley initialisation via polarized laser excitation makes ML MoS$_2$ an extremely promising system for valleytronics and investigating the robustness of the valley degree of freedom is of key importance for a practical implementation.
Our approach is to trace the change of valley index experimentally via the photon emission, that would have the opposite polarization compared to the excitation laser \cite{spinpol}. The robustness of the optically created polarization is confirmed in strong transverse magnetic fields and at temperatures up to 300K. Our experimental work allows to further test the association of the high PL emission polarization with selective valley excitation. \\
\indent \textit{Samples and Set-up.---} MoS$_2$ flakes are obtained by micro-mechanical cleavage of a natural bulk MoS$_2$ crystal \cite{Novoselov:2005a}(from SPI Supplies, USA) on a Si/90 nm SiO$_2$ substrate. The ML region is identified by optical contrast and very clearly in PL spectroscopy, based on the fact that MoS$_2$ becomes a direct seminconductor only for  a thickness $<2$~ML. Experiments on the same sample as in Ref.\cite{Cao:2012a} are carried out in a home build confocal microscope optimized for polarized PL experiments\cite{Sallen:2011a} in an attoDry magneto-cryostat. The detection spot diameter is $\approx1$~$\mu$m, optical excitation is achieved with a HeNe laser ($1.96$~eV, typical power $50 \mu W$) that is for $T<50$~K close the resonance with the lowest energy, direct transition (red transition $K_+$ valley in Fig.\ref{fig:fig2}a)  and the resulting emission polarization is analysed in the circular basis, dispersed in a spectrometer and detected with an Si-CCD camera. For higher energy excitation a frequency doubled Nd:YAG laser at 2.33~eV is used (as in Fig.\ref{fig:fig2}b). The sample temperature can be controlled between 4 and 300K, a transverse magnetic field $B_T$ up to 9~T in the plane of the MoS$_2$ monolayer  (Voigt geometry) is applied, see inset of Fig.\ref{fig:fig3}a.\\
\indent \textit{Experimental Results.---} First we investigate the temperature dependence of the PL polarization, monitoring if the optically created carriers can change valley in k-space, for example through scattering with phonons of suitable wavevector. Before discussing polarization, we note that the overall spectral and temperature dependence of the PL emission is very similar to Ref. \cite{Korn:2011a} where the same substrate was used: independent of the laser excitation power, we observe a low energy peak \textbf{\textit{L}} centred at 4K close to  $1.8$~eV, which is only detectable for $T<175$~K. The exact origin of peak \textbf{\textit{L}}, most likely linked to bound exciton states \cite{Korn:2011a}, is still under investigation. The emission intensity of peak \textbf{\textit{L}} has been found to depend on the substrate material used, SiO$_2$ or BN \cite{Mak:2012a}, and can be strongly reduced through oxide coverage \cite{Plechinger:2012a}. The reminder of the discussion is devoted to the high energy peak \textbf{\textit{A}} observed from 4~K (centred at $\approx 1.9$~eV) to room temperature which is attributed to the free exciton emission, expected to obey the selection rules presented in Fig.\ref{fig:fig2}a.\\
\indent We can separately access the polarization properties of the two dominant peaks \textbf{\textit{A}} and \textbf{\textit{L}} of the emission by fitting two Gaussians as in Fig.\ref{fig:fig1}c. Following excitation with a circularly $\sigma^+$ polarized HeNe laser we observe at $T=4$~K strongly circularly $\sigma^+$ polarized emission, confirming very recent reports \cite{Mak:2012a,Zeng:2012a,Cao:2012a}. Here the polarization of the free exciton peak \textbf{\textit{A}} is about 90\%, see Fig.\ref{fig:fig1}b, close to the theoretically predicted 100\% for the direct transition. 
When raising the temperature no measurable change occurs for the \textbf{\textit{A}} line polarization up to about $T=50$~K, beyond this temperature a steady decrease of the emitted polarization is observed.  At room temperature, a substantial polarization of 40\% remains, as demonstrated in Fig.\ref{fig:fig1}c, indicating that selective optical k-valley excitation with a polarized laser is still possible.\\
Fig.\ref{fig:fig2}b presents the $\sigma^+$ and $\sigma^-$ circularly polarized PL intensity and the corresponding circular polarization following highly, non-resonant excitation at 2.33~eV. In this case both \textbf{\textit{A}}  (as observed in Fig.\ref{fig:fig1}) and \textbf{\textit{B}} VB to CB transitions are possible and we observe indeed hot luminescence from the CB to \textbf{\textit{B}} band transition, which is polarized up to about 30\%. At this high laser energy the valley selectivity is not very high as the ground state transition \textit{A} is only about 10\% polarized. The distance in energy between \textbf{\textit{A}} and \textbf{\textit{B}} bands is  $\approx170$~meV, in good agreement with the predicted 150~meV \cite{Zhu:2011a}. Based on the predicted, strict valley selectivity, hot luminescence from the CB to \textbf{\textit{B}} transition is polarized dominantly $\sigma^+$, because the radiative recombination rate is fast compared to the combined rates of the processes necessary to emit $\sigma^-$ polarized light (i.e. change in K-valley, spin and orbital angular momentum) \cite{bpol}.


\indent A second crucial experimental test for the stability of the optically created valley polarization is the application of a transverse magnetic field. The spin quantization axis is chosen along the light propagation direction, i.e. perpendicular to the MoS$_2$ layer as indicated in the inset of Fig.\ref{fig:fig3}a, where the magnetic field B$_T$ is applied in the MoS$_2$ plane. In a classical picture for free carrier spins, the application of a magnetic field perpendicular to the spin quantization axis results in a precession of the spin around the applied field. As a result the spin component along the quantization axis measured via the PL  polarization degree in the circular basis would strongly decrease and eventually average out to zero in standard Hanle experiments \cite{Dyakonov:2008a}. With this simple picture in mind the constant polarization observed at $T=4$~K in transverse magnetic fields up to 9T  and at $T=100$~K in a field of $B_T=6$~T shown in Fig.\ref{fig:fig3}b seems very surprising. Again one needs to take into account the very unusual MoS$_2$ bandstructure, as discussed below.\\
\indent \textit{Discussion.---} Both series of experiments detailed in Fig.\ref{fig:fig1} and Fig.\ref{fig:fig3} show robust PL polarization following $\sigma^+$ polarized laser excitation i.e. the processes leading to a change in valley state for both conduction and valence carriers and hence $\sigma^-$ polarized emission are not very efficient. 
The emission is based on the dipole allowed recombination of a conduction electron with a valence hole, that we have detailed in Fig.\,\ref{fig:fig2}a. The main contribution to the strong optical absorption reported of nearly 10\% \cite{Mak:2010a,Mak:2012a,Splendiani:2010a,Eda:2011a} comes from the Mo $d$ orbitals (characterized by $\alpha$), taking into account the contribution of transitions involving states in adjacent cells in k-space. $\beta$ characterizes the minute contribution of the $p$ orbitals ($L=1$) to the $p-d$ hybridized VB states \cite{oscB}. The local trigonal symmetry of electron states in monolayer MoS$_2$ at $K_{\pm}$ points is $C_{3h}$ in Koster notation \cite{Koster:1963a}, and we have accordingly labelled the states using the corresponding group representations to which they belong. This leads us to predict that: (i) the only symmetry allowed coupling introduced by a transverse magnetic field $B_T$ is in principle between VB \textbf{\textit{A}} and VB \textbf{\textit{B}}, and between VB \textbf{\textit{A}} and CB, all of the same valley. This coupling would \textit{not} change the emitted light polarization. In addition, the large energy difference between these bands makes this coupling vanishingly small. (ii) $B_T$ does not couple states of inequivalent $K_+$ and $K_-$ valleys.
If the states have indeed the predicted symmetry, even a strong transverse field $B_T$ would result in a constant emission polarization when increasing $B_T$. Our results seem to validate this prediction, even at elevated temperatures of 100~K in a field of $B_T=6$~T as shown in Fig.\ref{fig:fig3}b, in addition to the initial reports in \cite{Zeng:2012a} for small fields. \\
Note that we do not have a clear spectral signature that would allow identifying the exciton charge state as suggested in Ref.\cite{Mak:2012a}. We therefore assume, in accordance with the intriguing polarization properties, the existence of neutral X$^0$ and/or negatively charged excitons X$^-$. The negatively charged exciton X$^-$ consists of two electrons in a singlet state and an unpaired hole. Measuring the PL polarization allows monitoring directly the orientation of the hole spin as electron-hole Coulomb exchange effects essentially cancel out. For the X$^0$, following $\sigma^{\pm}$ photon absorption, bright excitons with a total (pseudo-)spin of +1 or -1 are created. Dipole allowed optical transitions conserve the spin. So if only one of the carriers (electron or hole) flips, the X$^0$ becomes dark and non-radiative recombination might be favoured. An electron spin flip is energetically feasible based on (nearly) degenerate conduction states. Hole spin flips, at the centre of the investigation here, involve changing valley and are hence far less likely. To observe PL polarization change from the recombination of an exciton both carriers have to be flipped (the exciton spin flips as a whole) \cite{Dyakonov:2008a}. Importantly, to observe $\sigma^-$ emission following optical excitation with $\sigma^+$ light, involves for both X$^0$ and the X$^-$ a hole spin flip by going from valley $K_+$ to $K_-$. In this context, changing the temperature has several consequences:\\
\textbf{(i)} Phonon assisted intervalley scattering ($K_+\rightarrow K_-$) can be activated, since the average occupation number of acoustic and optical phonons increases for modes with wavevector $q\simeq |K_+-K_-|$, thus increasing the probability of intervalley transfer for both electrons and holes.\\
\textbf{(ii)} As the temperature is increased, the band gap decreases \cite{Korn:2011a} but the laser energy in Fig.\ref{fig:fig1} is kept constant at 1.96~eV, which is very close to resonance with the top VB \textbf{\textit{A}} to bottom of the CB direct transition at $T<50$~K. In a simple, independent particle picture, when raising the temperature, carriers further and further away from the $K_+$ point are optically injected \cite{footEXCITON} and the consistently high PL polarization observed in Fig.\ref{fig:fig1}a indicates that the chiral valley selectivity is as predicted very high throughout the valley \cite{Cao:2012a}. At room temperature the laser is close to resonance with VB \textbf{\textit{B}} with the opposite spin in the same valley. Note that this does not change the emission polarization (still $\sigma^+$) if the $K_+$ valley is populated which could explain why we still measure 40\% polarization at 300~K. To fully understand the origin of the remaining PL polarization, not observed to this degree in previous reports \cite{Mak:2012a,Zeng:2012a}, time resolved measurements should be employed in the future to extract the exact ratio of the radiative lifetime as compared to the valley/spin lifetime that determines the steady state PL polarization.  Also, the fidelity of valley initialisation at 300~K could be strongly improved in future experiments using a laser energy closer to the \textbf{\textit{A}} to CB resonance. We record a higher polarization at elevated temperature than the MoS$_2$ ML on BN investigated in Ref. \cite{Mak:2012a}, indicating that the substrate material influences the opto-electronic properties of atomically flat MoS$_2$.\\
\textbf{(iii)} Due to the increase of the electron average kinetic energy, the average k-linear terms responsible for the electron spin splitting increase, while the elastic collision time decreases within a given valley. This may lead to an overall decrease of the conduction electron-spin relaxation time \cite{Dyakonov:2008a}.  However, we re-emphasize that an inter valley transfer both for electrons \textit{and} holes is necessary to observe PL polarization decay. \\
\indent The exact role of the strong Coulomb interaction also deserves further attention. Strong excitonic effects in this system close to the ideal two-dimensional confinement limit have been predicted \cite{Tawinan:2012a,Olson:2011a} with exciton binding energies of $E_{Ryd}^{2D}\approx800$~meV (Bohr radius $a_B\approx1$nm), which raises even questions concerning the most reliable determination of direct bandgap of MoS$_2$. \\
\indent \textit{Conclusion.---} The robust polarization observed at elevated temperatures and in transverse magnetic fields up to 9T is a strong indication that optically created carriers remain during their radiative lifetime in the k-space valley selected via the excitation laser helicity, in agreement with the very recently predicted coupled spin and valley physics in monolayers of MoS$_2$ \cite{Xiao:2012a,Cao:2012a,Zhu:2011a}. Also, as changing spin state means changing valley in k-space for valence states, the spin states are as the valley index very robust, which makes these results very interesting for future valley Hall coupled to spin Hall measurements and should encourage further research in this direction \cite{Li:2012a}.\\
\indent \textit{Acknowledgements.---} We thank France-China NSFC-ANR research project SPINMAN, ANR QUAMOS and ITN SPINOPTRONICS. XM acknowledges a Chinese Academy of Science visiting professorship for senior international scientists Gr.No. 2011T1J37. We thank Iann Gerber and Sergej Kunz for technical support.

%\bibliography{mos2prbNEW}% Produces the bibliography via BibTeX.

\begin{thebibliography}{29}
\expandafter\ifx\csname natexlab\endcsname\relax\def\natexlab#1{#1}\fi
\expandafter\ifx\csname bibnamefont\endcsname\relax
  \def\bibnamefont#1{#1}\fi
\expandafter\ifx\csname bibfnamefont\endcsname\relax
  \def\bibfnamefont#1{#1}\fi
\expandafter\ifx\csname citenamefont\endcsname\relax
  \def\citenamefont#1{#1}\fi
\expandafter\ifx\csname url\endcsname\relax
  \def\url#1{\texttt{#1}}\fi
\expandafter\ifx\csname urlprefix\endcsname\relax\def\urlprefix{URL }\fi
\providecommand{\bibinfo}[2]{#2}
\providecommand{\eprint}[2][]{\url{#2}}

\bibitem[{\citenamefont{Novoselov et~al.}(2004)\citenamefont{Novoselov, Geim,
  Morozov, Jiang, Zhang, Dubonos, Grigorieva, and Firsov}}]{Novoselov:2004a}
\bibinfo{author}{\bibfnamefont{K.~S.} \bibnamefont{Novoselov}},
  \bibinfo{author}{\bibfnamefont{A.~K.} \bibnamefont{Geim}},
  \bibinfo{author}{\bibfnamefont{S.}~\bibnamefont{Morozov}},
  \bibinfo{author}{\bibfnamefont{D.}~\bibnamefont{Jiang}},
  \bibinfo{author}{\bibfnamefont{Y.}~\bibnamefont{Zhang}},
  \bibinfo{author}{\bibfnamefont{S.}~\bibnamefont{Dubonos}},
  \bibinfo{author}{\bibfnamefont{I.}~\bibnamefont{Grigorieva}},
  \bibnamefont{and} \bibinfo{author}{\bibfnamefont{A.~A.}
  \bibnamefont{Firsov}}, \bibinfo{journal}{Science}
  \textbf{\bibinfo{volume}{306}}, \bibinfo{pages}{666} (\bibinfo{year}{2004}).

\bibitem[{\citenamefont{Castro~Neto et~al.}(2009)\citenamefont{Castro~Neto,
  Guinea, Peres, Novoselov, and Geim}}]{Castroneto:2009a}
\bibinfo{author}{\bibfnamefont{A.~H.} \bibnamefont{Castro~Neto}},
  \bibinfo{author}{\bibfnamefont{F.}~\bibnamefont{Guinea}},
  \bibinfo{author}{\bibfnamefont{N.~M.~R.} \bibnamefont{Peres}},
  \bibinfo{author}{\bibfnamefont{K.~S.} \bibnamefont{Novoselov}},
  \bibnamefont{and} \bibinfo{author}{\bibfnamefont{A.~K.} \bibnamefont{Geim}},
  \bibinfo{journal}{Rev. Mod. Phys.} \textbf{\bibinfo{volume}{81}},
  \bibinfo{pages}{109} (\bibinfo{year}{2009}).

\bibitem[{\citenamefont{Mak et~al.}(2010)\citenamefont{Mak, Lee, Hone, Shan,
  and Heinz}}]{Mak:2010a}
\bibinfo{author}{\bibfnamefont{K.~F.} \bibnamefont{Mak}},
  \bibinfo{author}{\bibfnamefont{C.}~\bibnamefont{Lee}},
  \bibinfo{author}{\bibfnamefont{J.}~\bibnamefont{Hone}},
  \bibinfo{author}{\bibfnamefont{J.}~\bibnamefont{Shan}}, \bibnamefont{and}
  \bibinfo{author}{\bibfnamefont{T.~F.} \bibnamefont{Heinz}},
  \bibinfo{journal}{Phys. Rev. Lett.} \textbf{\bibinfo{volume}{105}},
  \bibinfo{pages}{136805} (\bibinfo{year}{2010}).

\bibitem[{\citenamefont{Splendiani et~al.}(2010)\citenamefont{Splendiani, Sun,
  Zhang, Li, Kim, Chim, Galli, and Wang}}]{Splendiani:2010a}
\bibinfo{author}{\bibfnamefont{A.}~\bibnamefont{Splendiani}},
  \bibinfo{author}{\bibfnamefont{L.}~\bibnamefont{Sun}},
  \bibinfo{author}{\bibfnamefont{Y.}~\bibnamefont{Zhang}},
  \bibinfo{author}{\bibfnamefont{T.}~\bibnamefont{Li}},
  \bibinfo{author}{\bibfnamefont{J.}~\bibnamefont{Kim}},
  \bibinfo{author}{\bibfnamefont{C.-Y.} \bibnamefont{Chim}},
  \bibinfo{author}{\bibfnamefont{G.}~\bibnamefont{Galli}}, \bibnamefont{and}
  \bibinfo{author}{\bibfnamefont{F.}~\bibnamefont{Wang}},
  \bibinfo{journal}{Nano Letters} \textbf{\bibinfo{volume}{10}},
  \bibinfo{pages}{1271} (\bibinfo{year}{2010}).

\bibitem[{\citenamefont{Eda et~al.}(2011)\citenamefont{Eda, Yamaguchi, Voiry,
  Fujita, Chen, and Chhowalla}}]{Eda:2011a}
\bibinfo{author}{\bibfnamefont{G.}~\bibnamefont{Eda}},
  \bibinfo{author}{\bibfnamefont{H.}~\bibnamefont{Yamaguchi}},
  \bibinfo{author}{\bibfnamefont{D.}~\bibnamefont{Voiry}},
  \bibinfo{author}{\bibfnamefont{T.}~\bibnamefont{Fujita}},
  \bibinfo{author}{\bibfnamefont{M.}~\bibnamefont{Chen}}, \bibnamefont{and}
  \bibinfo{author}{\bibfnamefont{M.}~\bibnamefont{Chhowalla}},
  \bibinfo{journal}{Nano Letters} \textbf{\bibinfo{volume}{11}},
  \bibinfo{pages}{5111} (\bibinfo{year}{2011}).

\bibitem[{\citenamefont{Korn et~al.}(2011)\citenamefont{Korn, Heydrich, Hirmer,
  Schmutzler, and Sch\"{u}ller}}]{Korn:2011a}
\bibinfo{author}{\bibfnamefont{T.}~\bibnamefont{Korn}},
  \bibinfo{author}{\bibfnamefont{S.}~\bibnamefont{Heydrich}},
  \bibinfo{author}{\bibfnamefont{M.}~\bibnamefont{Hirmer}},
  \bibinfo{author}{\bibfnamefont{J.}~\bibnamefont{Schmutzler}},
  \bibnamefont{and}
  \bibinfo{author}{\bibfnamefont{C.}~\bibnamefont{Sch\"{u}ller}},
  \bibinfo{journal}{Applied Physics Letters} \textbf{\bibinfo{volume}{99}},
  \bibinfo{eid}{102109} (\bibinfo{year}{2011}).

\bibitem[{\citenamefont{Radisavljevic et~al.}(2011)\citenamefont{Radisavljevic,
  Radenovic, Brivio, Giacometti, and Kis}}]{Radisavljevic:2011a}
\bibinfo{author}{\bibfnamefont{B.}~\bibnamefont{Radisavljevic}},
  \bibinfo{author}{\bibfnamefont{A.}~\bibnamefont{Radenovic}},
  \bibinfo{author}{\bibfnamefont{J.}~\bibnamefont{Brivio}},
  \bibinfo{author}{\bibfnamefont{V.}~\bibnamefont{Giacometti}},
  \bibnamefont{and} \bibinfo{author}{\bibfnamefont{A.}~\bibnamefont{Kis}},
  \bibinfo{journal}{Nature. Nanotech.} \textbf{\bibinfo{volume}{6}},
  \bibinfo{pages}{147} (\bibinfo{year}{2011}).

\bibitem[{\citenamefont{Xiao et~al.}(2012)\citenamefont{Xiao, Liu, Feng, Xu,
  and Yao}}]{Xiao:2012a}
\bibinfo{author}{\bibfnamefont{D.}~\bibnamefont{Xiao}},
  \bibinfo{author}{\bibfnamefont{G.-B.} \bibnamefont{Liu}},
  \bibinfo{author}{\bibfnamefont{W.}~\bibnamefont{Feng}},
  \bibinfo{author}{\bibfnamefont{X.}~\bibnamefont{Xu}}, \bibnamefont{and}
  \bibinfo{author}{\bibfnamefont{W.}~\bibnamefont{Yao}},
  \bibinfo{journal}{Phys. Rev. Lett.} \textbf{\bibinfo{volume}{108}},
  \bibinfo{pages}{196802} (\bibinfo{year}{2012}).

\bibitem[{\citenamefont{Cao et~al.}(2012)\citenamefont{Cao, Wang, Han, Ye, Zhu,
  Shi, Niu, Tan, Wang, Liu et~al.}}]{Cao:2012a}
\bibinfo{author}{\bibfnamefont{T.}~\bibnamefont{Cao}},
  \bibinfo{author}{\bibfnamefont{G.}~\bibnamefont{Wang}},
  \bibinfo{author}{\bibfnamefont{W.}~\bibnamefont{Han}},
  \bibinfo{author}{\bibfnamefont{H.}~\bibnamefont{Ye}},
  \bibinfo{author}{\bibfnamefont{C.}~\bibnamefont{Zhu}},
  \bibinfo{author}{\bibfnamefont{J.}~\bibnamefont{Shi}},
  \bibinfo{author}{\bibfnamefont{Q.}~\bibnamefont{Niu}},
  \bibinfo{author}{\bibfnamefont{P.}~\bibnamefont{Tan}},
  \bibinfo{author}{\bibfnamefont{E.}~\bibnamefont{Wang}},
  \bibinfo{author}{\bibfnamefont{B.}~\bibnamefont{Liu}}, \bibnamefont{et~al.},
  \bibinfo{journal}{Nature Communications} \textbf{\bibinfo{volume}{3}},
  \bibinfo{pages}{887} (\bibinfo{year}{2012}).

\bibitem[{\citenamefont{Zhu et~al.}(2011)\citenamefont{Zhu, Cheng, and
  Schwingenschl\"ogl}}]{Zhu:2011a}
\bibinfo{author}{\bibfnamefont{Z.~Y.} \bibnamefont{Zhu}},
  \bibinfo{author}{\bibfnamefont{Y.~C.} \bibnamefont{Cheng}}, \bibnamefont{and}
  \bibinfo{author}{\bibfnamefont{U.}~\bibnamefont{Schwingenschl\"ogl}},
  \bibinfo{journal}{Phys. Rev. B} \textbf{\bibinfo{volume}{84}},
  \bibinfo{pages}{153402} (\bibinfo{year}{2011}).

\bibitem[{\citenamefont{Mak et~al.}(2012)\citenamefont{Mak, He, Shan, and
  Heinz}}]{Mak:2012a}
\bibinfo{author}{\bibfnamefont{K.~F.} \bibnamefont{Mak}},
  \bibinfo{author}{\bibfnamefont{K.}~\bibnamefont{He}},
  \bibinfo{author}{\bibfnamefont{J.}~\bibnamefont{Shan}}, \bibnamefont{and}
  \bibinfo{author}{\bibfnamefont{T.~F.} \bibnamefont{Heinz}},
  \bibinfo{journal}{Nature Nanotech.} \textbf{\bibinfo{volume}{DOI:
  10.1038/NNANO.2012.96}} (\bibinfo{year}{2012}).

\bibitem[{\citenamefont{Zeng et~al.}(2012)\citenamefont{Zeng, Dai, Yao, Xiao,
  and Cui}}]{Zeng:2012a}
\bibinfo{author}{\bibfnamefont{H.}~\bibnamefont{Zeng}},
  \bibinfo{author}{\bibfnamefont{J.}~\bibnamefont{Dai}},
  \bibinfo{author}{\bibfnamefont{W.}~\bibnamefont{Yao}},
  \bibinfo{author}{\bibfnamefont{D.}~\bibnamefont{Xiao}}, \bibnamefont{and}
  \bibinfo{author}{\bibfnamefont{X.}~\bibnamefont{Cui}},
  \bibinfo{journal}{Nature Nanotech.} \textbf{\bibinfo{volume}{DOI:
  10.1038/NNANO.2012.95}} (\bibinfo{year}{2012}).

\bibitem[{CB()}]{CB}
\bibinfo{note}{The conduction band spin splitting due to a k-linear term is
  small near $K_{\pm}$ compared to the VB spin splitting.}

\bibitem[{\citenamefont{Dyakonov}(2008)}]{Dyakonov:2008a}
\bibinfo{author}{\bibfnamefont{M.}~\bibnamefont{Dyakonov}},
  \bibinfo{journal}{Springer Series in Solid-State Science, Springer-Verlag
  Berlin} \textbf{\bibinfo{volume}{157}} (\bibinfo{year}{2008}).

\bibitem[{\citenamefont{Gunawan et~al.}(2006)\citenamefont{Gunawan, Shkolnikov,
  Vakili, Gokmen, De~Poortere, and Shayegan}}]{Gunawan:2006a}
\bibinfo{author}{\bibfnamefont{O.}~\bibnamefont{Gunawan}},
  \bibinfo{author}{\bibfnamefont{Y.~P.} \bibnamefont{Shkolnikov}},
  \bibinfo{author}{\bibfnamefont{K.}~\bibnamefont{Vakili}},
  \bibinfo{author}{\bibfnamefont{T.}~\bibnamefont{Gokmen}},
  \bibinfo{author}{\bibfnamefont{E.~P.} \bibnamefont{De~Poortere}},
  \bibnamefont{and} \bibinfo{author}{\bibfnamefont{M.}~\bibnamefont{Shayegan}},
  \bibinfo{journal}{Phys. Rev. Lett.} \textbf{\bibinfo{volume}{97}},
  \bibinfo{pages}{186404} (\bibinfo{year}{2006}).

\bibitem[{\citenamefont{Zhang et~al.}(2011)\citenamefont{Zhang, Jung, Fiete,
  Niu, and MacDonald}}]{Zhang:2011a}
\bibinfo{author}{\bibfnamefont{F.}~\bibnamefont{Zhang}},
  \bibinfo{author}{\bibfnamefont{J.}~\bibnamefont{Jung}},
  \bibinfo{author}{\bibfnamefont{G.~A.} \bibnamefont{Fiete}},
  \bibinfo{author}{\bibfnamefont{Q.}~\bibnamefont{Niu}}, \bibnamefont{and}
  \bibinfo{author}{\bibfnamefont{A.~H.} \bibnamefont{MacDonald}},
  \bibinfo{journal}{Phys. Rev. Lett.} \textbf{\bibinfo{volume}{106}},
  \bibinfo{pages}{156801} (\bibinfo{year}{2011}).

\bibitem[{\citenamefont{Rycerz et~al.}(2007)\citenamefont{Rycerz, Tworzydlo,
  and Beenakker}}]{Rycerz:2007a}
\bibinfo{author}{\bibfnamefont{A.}~\bibnamefont{Rycerz}},
  \bibinfo{author}{\bibfnamefont{J.}~\bibnamefont{Tworzydlo}},
  \bibnamefont{and} \bibinfo{author}{\bibfnamefont{C.~J.}
  \bibnamefont{Beenakker}}, \bibinfo{journal}{Nature. Phys.}
  \textbf{\bibinfo{volume}{3}}, \bibinfo{pages}{172} (\bibinfo{year}{2007}).

\bibitem[{\citenamefont{Zhou et~al.}(2007)\citenamefont{Zhou, Gweon, Fedorov,
  First, de~Heer, Lee, Guinea, Neto, and Lanzara}}]{Zhou:2007a}
\bibinfo{author}{\bibfnamefont{S.~Y.} \bibnamefont{Zhou}},
  \bibinfo{author}{\bibfnamefont{G.-H.} \bibnamefont{Gweon}},
  \bibinfo{author}{\bibfnamefont{A.~V.} \bibnamefont{Fedorov}},
  \bibinfo{author}{\bibfnamefont{P.~N.} \bibnamefont{First}},
  \bibinfo{author}{\bibfnamefont{W.~A.} \bibnamefont{de~Heer}},
  \bibinfo{author}{\bibfnamefont{D.-H.} \bibnamefont{Lee}},
  \bibinfo{author}{\bibfnamefont{F.}~\bibnamefont{Guinea}},
  \bibinfo{author}{\bibfnamefont{A.~H.~C.} \bibnamefont{Neto}},
  \bibnamefont{and} \bibinfo{author}{\bibfnamefont{A.}~\bibnamefont{Lanzara}},
  \bibinfo{journal}{Nature Mater.} \textbf{\bibinfo{volume}{6}},
  \bibinfo{pages}{770} (\bibinfo{year}{2007}).

\bibitem[{spi()}]{spinpol}
\bibinfo{note}{Measuring the polarization of the emitted photon itself is not
  sufficient to deduce the exact carrier spin state in a given valley, the
  transitions energy has to be taken into account i.e. it is possible to excite
  with $\sigma^+$ polarized light a pure spin state in the $K_+$ valley, if in
  resonance with the lowest energy transition (red arrow in
  Fig.\ref{fig:fig2}a).}

\bibitem[{\citenamefont{Novoselov et~al.}(2005)\citenamefont{Novoselov, Jiang,
  Schedin, Booth, Khotkevich, Morozov, and Geim}}]{Novoselov:2005a}
\bibinfo{author}{\bibfnamefont{K.~S.} \bibnamefont{Novoselov}},
  \bibinfo{author}{\bibfnamefont{D.}~\bibnamefont{Jiang}},
  \bibinfo{author}{\bibfnamefont{F.}~\bibnamefont{Schedin}},
  \bibinfo{author}{\bibfnamefont{T.~J.} \bibnamefont{Booth}},
  \bibinfo{author}{\bibfnamefont{V.~V.} \bibnamefont{Khotkevich}},
  \bibinfo{author}{\bibfnamefont{S.~V.} \bibnamefont{Morozov}},
  \bibnamefont{and} \bibinfo{author}{\bibfnamefont{A.~K.} \bibnamefont{Geim}},
  \bibinfo{journal}{Proc. Natl Acad. Sci. USA} \textbf{\bibinfo{volume}{102}},
  \bibinfo{pages}{10451} (\bibinfo{year}{2005}).

\bibitem[{\citenamefont{Sallen et~al.}(2011)\citenamefont{Sallen, Urbaszek,
  Glazov, Ivchenko, Kuroda, Mano, Kunz, Abbarchi, Sakoda, Lagarde
  et~al.}}]{Sallen:2011a}
\bibinfo{author}{\bibfnamefont{G.}~\bibnamefont{Sallen}},
  \bibinfo{author}{\bibfnamefont{B.}~\bibnamefont{Urbaszek}},
  \bibinfo{author}{\bibfnamefont{M.~M.} \bibnamefont{Glazov}},
  \bibinfo{author}{\bibfnamefont{E.~L.} \bibnamefont{Ivchenko}},
  \bibinfo{author}{\bibfnamefont{T.}~\bibnamefont{Kuroda}},
  \bibinfo{author}{\bibfnamefont{T.}~\bibnamefont{Mano}},
  \bibinfo{author}{\bibfnamefont{S.}~\bibnamefont{Kunz}},
  \bibinfo{author}{\bibfnamefont{M.}~\bibnamefont{Abbarchi}},
  \bibinfo{author}{\bibfnamefont{K.}~\bibnamefont{Sakoda}},
  \bibinfo{author}{\bibfnamefont{D.}~\bibnamefont{Lagarde}},
  \bibnamefont{et~al.}, \bibinfo{journal}{Phys. Rev. Lett.}
  \textbf{\bibinfo{volume}{107}}, \bibinfo{pages}{166604}
  (\bibinfo{year}{2011}).

\bibitem[{\citenamefont{Plechinger et~al.}(2012)\citenamefont{Plechinger,
  Schrettenbrunner, Eroms, Weiss, Sch\"uller, and Korn}}]{Plechinger:2012a}
\bibinfo{author}{\bibfnamefont{G.}~\bibnamefont{Plechinger}},
  \bibinfo{author}{\bibfnamefont{F.-X.} \bibnamefont{Schrettenbrunner}},
  \bibinfo{author}{\bibfnamefont{J.}~\bibnamefont{Eroms}},
  \bibinfo{author}{\bibfnamefont{D.}~\bibnamefont{Weiss}},
  \bibinfo{author}{\bibfnamefont{C.}~\bibnamefont{Sch\"uller}},
  \bibnamefont{and} \bibinfo{author}{\bibfnamefont{T.}~\bibnamefont{Korn}},
  \bibinfo{journal}{phys. stat. sol. (RRL)} \textbf{\bibinfo{volume}{6}},
  \bibinfo{pages}{126} (\bibinfo{year}{2012}).

\bibitem[{bpo()}]{bpol}
\bibinfo{note}{The polarization observed for peak \textbf{\textit{B}} is not
  100\% probably due to the fact that the 2.33~eV Laser excitation
  photogenerates carriers far from the minimum of the $K_+$ and $K_-$ valleys,
  where optical selection rules might be very different as compared to states
  close to the K$_{\pm}$ points.}

\bibitem[{Kos()}]{Koster:1963a}
\bibinfo{note}{G. F. Koster, J. O. Dimmock, G. Wheeler, R. G. Satz,
  \textit{Properties of thirty-two point groups} (M.I.T. Press, Cambridge,
  Massachusetts USA) 1963.}

\bibitem[{osc()}]{oscB}
\bibinfo{note}{The ratio $|\beta'/\alpha'|$ for VB \textbf{\textit{B}} has to
  the best of our knowledge not been reported in the literature up to now.}

\bibitem[{foo()}]{footEXCITON}
\bibinfo{note}{Due to the presence of strong excitonic effects the A and B
  resonances are composed of an ensemble of electronic states occupying a large
  portion of the Brillouin zone.}

\bibitem[{\citenamefont{Cheiwchanchamnangij and
  Lambrecht}(2012)}]{Tawinan:2012a}
\bibinfo{author}{\bibfnamefont{T.}~\bibnamefont{Cheiwchanchamnangij}}
  \bibnamefont{and} \bibinfo{author}{\bibfnamefont{W.~R.~L.}
  \bibnamefont{Lambrecht}}, \bibinfo{journal}{Phys. Rev. B}
  \textbf{\bibinfo{volume}{85}}, \bibinfo{pages}{205302}
  (\bibinfo{year}{2012}).

\bibitem[{\citenamefont{Olsen et~al.}(2011)\citenamefont{Olsen, Jacobsen, and
  Thygesen}}]{Olson:2011a}
\bibinfo{author}{\bibfnamefont{T.}~\bibnamefont{Olsen}},
  \bibinfo{author}{\bibfnamefont{K.~W.} \bibnamefont{Jacobsen}},
  \bibnamefont{and} \bibinfo{author}{\bibfnamefont{K.~S.}
  \bibnamefont{Thygesen}}, \bibinfo{journal}{e-print}
  \textbf{\bibinfo{volume}{arXiv:1107.0600}} (\bibinfo{year}{2011}).

\bibitem[{\citenamefont{Li et~al.}(2012)\citenamefont{Li, Zhang, and
  Niu}}]{Li:2012a}
\bibinfo{author}{\bibfnamefont{X.}~\bibnamefont{Li}},
  \bibinfo{author}{\bibfnamefont{F.}~\bibnamefont{Zhang}}, \bibnamefont{and}
  \bibinfo{author}{\bibfnamefont{Q.}~\bibnamefont{Niu}},
  \bibinfo{journal}{e-print} \textbf{\bibinfo{volume}{arXiv:1207.1205}}
  (\bibinfo{year}{2012}).

\end{thebibliography}

\end{document}